2022-08-08

# Getting around the Halting Problem

*by X.Y. Newberry*

**Abstract**

There are numbers *k* and *s* and a URM program *A(n,m)* satisfying the following conditions.

1. If *A(n,m)* halts, then $C_n(m)$ diverges.
2. For all *n*, $C_k(n) = A(n,n)$ and $C_s(n) = C_k(s)$.
3. *A(k,s)* halts and for all *n*, *A(s,n)* diverges.

Here $C_n(_)$ is a program with index *n* in some exhaustive enumeration of all possible programs.

This has implications for solving the liar paradox and for generalization of Gödel's incompleteness theorem to formal systems other than PA.

# 1. Introduction

Let ***A***(n,m) be a program[1]) with the following property[2])

$$\boldsymbol{A}(n,m)\downarrow \rightarrow C_n(m)\uparrow \qquad (1.1)$$

that is, ***A***() detects in some cases that a program $C_n()$ with input *m* does not halt. Here $C_n()$ is a program with index *n* in some exhaustive enumeration of all possible programs. The downward arrow '↓' indicates that the program halts, the upward arrow '↑' indicates that the program diverges.

Let us define $C_k(n) =_{def} A(n,n)$, $C_s() =_{def} C_k(s)$. The existence of $C_s()$ is guaranteed by the Recursion Theorem. In pseudocode it looks as follows

```
C_k(n) {
    A(n,n)
}
```

Figure 1.1

and

```
C_s() {
      C_k(s)
}
```

Figure 1.2

It is apparent that C_s() can be constructed in any general purpose programming language. We know by the Halting Theorem that there is no program ***A()*** such that

$$\boldsymbol{A}(n,m)\downarrow \Longleftrightarrow C_n(m)\uparrow \qquad (1.2)$$

1 Such as URM program or a Turing machine performing operations on natural numbers. It could be a program in any general purpose programming language with the proviso that word length and memory size are infinite.

2 The idea of such a program *A()* is due to Penrose [5]. This approach greatly simplifies matters.

The proof is usually by constructing a counterexample:

(1) $C_s()\downarrow$ — Assumption 1
(2) $C_k(s)\downarrow$ — From 1, figure 1.1
(3) $A(s,s)\downarrow$ — From 2, figure 1.2
(4) $C_s()\uparrow$ — From 3, condition 1.1
**(5)** $\boldsymbol{C_s()\uparrow}$ — RAA 1, 4, assumption 1 discharged

Proof 1.1

But $C_s()\downarrow = C_k(s)\downarrow = A(s,_)\downarrow$ are all equivalent. So we can analogically prove $C_k(s)\uparrow$ and $A(s,_)\uparrow$. So $C_s()$ does not halt, but ***A****()* does not know it.

The question arises if there is any other way ***A***(n,m) can express that $C_s()$ does not halt.[3]) It is a thesis of this paper that there is. In particular it is the program *A(n,m)* such that

a) $A(\mathbf{k},s)\downarrow$
b) $A(\mathbf{s},_)\uparrow$

(The underscore in place of a parameter indicates 'don't care'.) The condition a) together with (1.1) imply that $C_k(s)\uparrow$. But if $C_k(s)$ does not halt then $C_s()$ cannot possibly halt either. (Figure 1.2) Thus $A(\mathbf{k},s)\downarrow$ expresses indirectly that $C_s()$ does not halt. Seemingly we have a contradiction: $C_s() = C_k(s)$ and yet $A(\mathbf{k},s)\downarrow$ & $A(\mathbf{s},_)\uparrow$. The key observation here is that $A(\mathbf{k},s)\downarrow$ does *not* contradict $A(\mathbf{s},_)\uparrow$. The former 'says' that $C_k(s)$ does not halt, the latter says nothing; it does ***not*** say that $C_s()$ *does* halt. And since <*k,s*> and <*s,_*> are two different parameter tuples *A()* is under no obligation to react identically to them.

## 2. Formalization

**Theorem 1:** There are numbers *k* and *s* and a program *A(n,m)* satisfying the following conditions.

1. If $A(n,m)\downarrow$, then $C_n(m)\uparrow$.
2. For all n, $C_k(n) = A(n,n)$ and $C_s(n) = C_k(s)$.
3. $A(k,s)\downarrow$ and for all n, $A(s,n)\uparrow$.

Here $C_n(\bullet)$ is a program with index *n* in some exhaustive enumeration of all possible programs.

---

3 Never say never!

**Proof:** Let $K$ denote the halting problem, that is, $K = \{e : C_e(e) \downarrow\}$. It is known that the complement of $K$, $\overline{K}$, is productive and therefore contains an infinite recursively enumerable (r.e.) subset. Let $f$ be a recursive function such that $f(0), f(1), f(2), \ldots$ is a one-one enumeration of an infinite r.e. subset of $\overline{K}$.

Define a partial-recursive function $A(x,y)$, and numbers $k$, $s$ such that for all $n$

$$A(x,y) = \begin{cases} 1 & \text{if } \exists n(x = y = f(n)) \vee (x = k \,\&\, y = s); \\ \uparrow & \text{otherwise.} \end{cases}$$

$$C_k(x) = \begin{cases} 1 & \text{if } A(x,x) = 1 \\ \uparrow & \text{otherwise.} \end{cases}$$

By the Recursion Theorem, there are infinitely many $s'$ such that for all $y$, $C_{s'}(y) = C_k(s')$. Fix one such $s'$, say $s$, such that $s \neq k$. Choose $f(n)$ with the property that no $k'$ such that $C_{k'} = C_k$ and no $s'$ such that no $C_{s'}(y) = C_k(s')$ are in its range. The function $f$ enumerates many functions that do *not* halt on their own input (hopefully all of them except the equivalents of $k$ and $s$.)

The definition is seemingly circular: $A$ refers to $C_k$ while $C_k$ is defined in terms of $A$. In fact it is a recursive definition: by the Recursion Theorem $A$ obtains its own description from which it derives the description of $C_k$.

We may now verify Conditions 1, 2 and 3 as follows.

Condition 1: By the definition of $A$, $A(x,y)\downarrow$ only if one of the following holds: (a) $x = y = f(n)$, or (b) $(x = k \,\&\, y = s)$. In case (a), one has by the definition of $f$ that $f(n) \in \overline{K}$, that is, $C_{f(n)}(f(n))\uparrow$. In case (b), since s has been chosen such that $s \neq k$, by the definition of $C_k(x)$, $C_k(x)\uparrow$ if $x \neq y$.

Condition 2: For all $n$, $C_k(n) = A(n,n)$ is immediate by the definition of $C_k$. Furthermore, by the choice of $s$, for all $y$, $C_s(y) = C_k(s)$.

Condition 3: Since $f$ was chosen such that for all $n$, $s \neq f(n)$, it follows from the definition of $A$ that $A(k,s)\downarrow = 1$ and for all $y$, $A(s,y)\uparrow$. QED [4])

4 This is a modified version of a proof suggested by a referee.

There is a compilable, executable version of another program in Appendix A written in C, which satisfies the same conditions as *A()* above. It perhaps constitutes an alternative proof.

# 3. Arithmetic

I have suggested in [3] that the Diagonal Lemma fails in Aristotelian logic/Strawson's logic of presuppositions [6]. This paper is the computability version of said argument. Consider Gödel's sentence

$$\sim\exists x\exists z(\mathrm{Prf}(x,z)\ \&\ \mathrm{Diag}(k,z)) \qquad \text{(G)}$$

where $\ulcorner G \urcorner$ is the Gödel number of G, and *Diag(k,z)* is satisfied only by $z = \ulcorner G \urcorner$. In classical logic

$$\sim(\exists x)Prf(x, \ulcorner G \urcorner) \qquad \text{(H')}$$

is equivalent to (G). But in Strawson's logic of presuppositions

$$(\exists x)Prf(x, \ulcorner G \urcorner) \qquad \text{(H)}$$

is a *presupposition* of (G). Hence when (H) is false then (G) is *neither true nor false*. If we could devise a non-classical logic where vacuous sentences are 'outlawed' then we could potentially derive H' but not G. (Such a logic has been proposed in [4].)

While (G) refers to itself, and 'says' "I am not provable", (H) says "(G) *is not provable*." We observe an analogy:

| | Computability | Logic |
|---|---|---|
| Self-reference | $C_s()$ | G |
| External reference | $C_k(s)$ | $\sim\exists x(\mathrm{Prf}(x, \ulcorner G \urcorner)$ |

Table 3.1

Program *A()* does *not* explicitly 'know' that $C_s()$ does not halt, but *does know* that $C_k(s)$ does not halt. And while Strawson based arithmetic does not 'know' that (G) is the case it nevertheless does know that (H') is the case. But all the information we are interested in is already contained in (H').

# 4. The Liar Paradox

Consider The following two-line puzzle:

**Line 1:** This sentence is not true.
**Line 2:** "This sentence is not true" is not true.

The standard evaluation rule for a sentence of the form 'The sentence X is true' is roughly this:

*(*) Go to the sentence X and evaluate it. If that sentence is true, so is 'The sentence X is true' , else the latter is false.*

When we apply the above rule to line 1 we end up in an infinite loop. The sentence cannot be evaluated, and hence is not true (and not false.) When we apply it to line 2 we already know that "This sentence is not true" is not true, and therefore the line 2 sentence *is* true. The "go-to" command makes the referring by 'The sentence ...' operationally explicit.[5])

Another way of putting it is this: *Y = This sentence is not true*.
*Y* is not true
*~T(Y)* is true
Although *Y* and *~T(Y)* appear to say the same thing, they do not. The former is self-referential, the latter is not. We observe an analogy

| | Computability | Logic | Liar |
|---|---|---|---|
| Self-reference | $C_s()$ | G | Y |
| External reference | $C_k(s)$ | ~∃x(Prf(x, ⌜G⌝) | ~T(Y) |

Table 4.1

More about this in [2].

5 All the above is a paraphrase of [1].

# APPENDIX A: A() fragment

A reasonable strategy for *A()* to determine if *C_s()* does not halt is to call *C_s()* itself. For *C_s()* is a program that is supposed to halt if *C_s()* does not halt. So A() can as well ask it. And after all *C_s() = C_k(s) = A(s,s)* are computationally equivalent. In this case when *A()* is called with s it will go into infinite recursion. The program below will actually compile and execute.

```c
#include <stdio.h>
#define DEPTH (1000)
#define INITIALIZE  level = 0; k = (unsigned int)C_k; \
    s = (unsigned int)C_s;

static int level = 0;
unsigned int s, k;

A(unsigned int n, unsigned int m);    // Forward declaration

int checkStack(void) {
    level++;
    if (level >= DEPTH){
        printf("Almost got there. Have run out of stack.\n\n");
        level--;
        return 1;
    }
    return 0;
}

C_k(unsigned int n) {
    A(n,n);
}

C_s() {
    C_k(s);
}

// If A(n,m) halts then C_n(m) does NOT halt.
A(unsigned int n, unsigned int m) {
    if ( checkStack() ) return;
    // . . . . .
    if (n == s) C_s();
    //  . . . . .
    if (n == k && m == s) {
        // Analyze what happens when A() calls C_s()
        // Does C_s() have a base case?
        printf("Program C_%u( %u ) does NOT halt.\n\n", n, m);
    }
    // . . . . .
    level--;
}

int main(int argc, char *argv[])
{
    INITIALIZE       // macro

    A(k,s);          // Does C_k(s) halt? No.
    C_k(s);          // Does C_s() halt? No answer.
    A(s,0);          // Does C_s() halt? No answer.

    return 0;
}
```

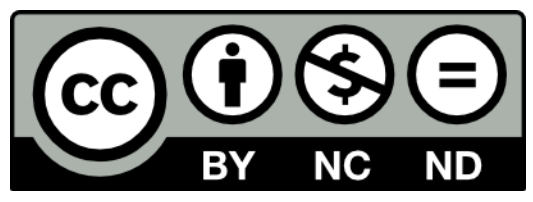